\documentclass[a4paper,10pt]{article}

\usepackage{graphicx}
\usepackage{caption}
\usepackage{subcaption}
\usepackage{moreverb}
\usepackage{bm}
\usepackage{dsfont}
\usepackage{fullpage}

\def\boxit#1{\vbox{\hrule\hbox{\vrule\kern6pt
          \vbox{\kern6pt#1\kern6pt}\kern6pt\vrule}\hrule}}

\begin{document}

\title{Lasso tuning through the flexible-weighted bootstrap}

\author{Ellis Patrick$^{*}$ and Samuel Mueller \\ School of Mathematics and Statistics, University of Sydney}








\maketitle

\abstract{
Regularized regression approaches such as the Lasso have been widely adopted for constructing sparse linear models in high-dimensional datasets.
A complexity in fitting these models is the tuning of the parameters which control the level of introduced sparsity through penalization.
The most common approach to select the penalty parameter is through $k$-fold cross-validation. 
While cross-validation is used to minimise the empirical prediction error, approaches such as the $m$-out-of-$n$ paired bootstrap which use smaller training datasets provide consistency in selecting the non-zero coefficients in the oracle model, performing well in an asymptotic setting but having limitations when $n$ is small.
In fact, for models such as the Lasso there is a monotonic relationship between the size of training sets and the penalty parameter.
We propose a generalization of these methods for selecting the regularization parameter based on a flexible-weighted bootstrap procedure that mimics the $m$-out-of-$n$ bootstrap and overcomes its challenges for all sample sizes.
Through simulation studies we demonstrate that when selecting a penalty parameter, the choice of weights in the bootstrap procedure can be used to dictate the size of the penalty parameter and hence the sparsity of the fitted model. 
We empirically illustrate our weighted bootstrap procedure by applying the Lasso to integrate clinical and microRNA data in the modeling of Alzheimer's disease.
In both the real and simulated data we find a narrow part of the parameter space to perform well, emulating an $m$-out-of-$n$ bootstrap, and that our procedure can be used to improve interpretation of other optimization heuristics.

}

\bigskip



\footnotetext{\textbf{Abbreviations:} CV, cross-validation; *\textbf{Corresponding author:} ellis.patrick@sydney.edu.au}

\section{Introduction}






The weighted bootstrap, where weights are repeatedly generated from the standard exponential distribution is one of the preferred ways in learning more about the stability of features for the best prediction of outcomes such as disease status or any other variable of interest, for example see Murray et al. \cite{murray_graphical_2013}, Barbe et al. \cite{barbe_weighted_1995}   and Minnier et al. \cite{minnier_perturbation_2011}. 
Using a weighted bootstrap is also a preferred method when dealing with multi-level or cluster longitudinal data \cite{chen_model_2018, oshaughnessy_bootstrapping_2018}.
Most recently, Tarr et al. \cite{tarr_package_2018} used this weighted bootstrap for generating a suite of visualisation tools for the selection of variables, including for learning from the Lasso.
%
 


The ubiquitous Lasso \cite{tibshirani_regression_1996} is a method of regression that can be applied in the high dimensional setting, even when $p > n$, to estimate the relationship  between a $p$-dimensional vector  of covariates  $x$  and  a response variable  $y$.
The Lasso is an L1 regularized version of ordinary least squares, 
\begin{equation}
\label{eq1}
\hat{\bm{\beta}} (\lambda) = \mathrm{arg\,min}_{{\bm{\beta}}} \left\{\sum\limits_{i=1}^n \left(y_i - \sum\limits_{j=1}^p \beta_j x_{ij}\right)^2 + \lambda \sum\limits_{j=1}^p |\beta_j| \right\},
\end{equation}
where $\lambda$ is a non-negative regularization parameter which controls the penalty imposed on the Lasso solution. If not otherwise made explicit, we assume that all the variables are standardized to have mean zero and sample variance one.
Convexity of the Lasso's L1 regularization enables efficient computation of the solution and several efficient algorithms have been developed, solving the Lasso along its solution path \cite{efron_least_2004, friedman_regularization_2010}.

The geometry of the Lasso's L1 regularization is such that greater values of the regularization parameter $\lambda$ will tend to result in greater shrinkage of the estimated regression coefficients, with the possibility that some coefficients are estimated as exactly zero for sufficiently large values of $\lambda$. 
Intrinsic to the Lasso is hence the assumption of sparsity, facilitating tractable application and interpretable solutions in the high dimensional setting. 
Tracking the non-zero parameter estimates along the solution path is critical for using the Lasso as a feature selection method and we define this so-called active set as $\hat{\alpha}(\lambda)= \{j : \hat\beta_j(\lambda) \ne 0; j=1,\ldots, p\}$, providing for each value of the tuning parameter $\lambda$ the indices of the corresponding regression coefficients that were not shrunk all the way to zero.
Along the Lasso solution path, great variation may exist between the active sets of the different Lasso models obtained through resampling.
This gives rise to the possibility of having very different interpretations of the underlying dataset, as well as very different predictive outcomes.
A crucial aspect of Lasso regression and one that is supported by much empirical and theoretical work \cite{yu_modified_2014, knight_asymptotics_2000, zou_adaptive_2006} is hence the selection of the Lasso regularization parameter.
In this article, we set as our primary objective to choose the Lasso regularization parameter, $\lambda$, that uncovers the true set of relevant covariates, assuming that such a set exists.


The methods for regularization parameter selection may largely be classified under the two broad categories of information criteria and resampling, although some close connections between the two can be made \cite{mueller_outlier_2005, mueller_model_2013}.
Whilst information criteria type methods such as the BIC \cite{schwarz_estimating_1978} and EBIC \cite{chen_extended_2008} are computationally cheaper, resampling procedures such as cross-validation and the bootstrap have the advantage of avoiding any assumptions about the structure and distribution of the underlying model. 
This allows resampling procedures to be applied to arbitrarily complicated situations including those for which parametric modeling and or theoretical analysis may be difficult or intractable \cite{efron_jackknife_1982}.

In practice, perhaps the most common means of selecting the Lasso regularization parameter is through $k$-fold cross-validation \cite{tibshirani_regression_1996,friedman_regularization_2010}.
This method, particularly when $k$ relatively to $n$ is large, is however both theoretically and empirically known to select models with too many covariates \cite{martinez_empirical_2011, yu_modified_2014, zhang_cross-validation_2015, su_when_2018}.
In a particular attempt to compensate for this, Friedman et al. \cite{friedman_regularization_2010} in their R package \texttt{glmnet} have suggested an empirical correction which is to choose the largest regularization parameter $\lambda$ that is within one standard error of the minimum mean cross-validated error. 
This has the effect of selecting a Lasso model that is more heavily penalised and hence more sparse than the model selected by the usual cross-validation procedure.
Another effort to overcome this over-selection problem was made by Feng and Yu \cite{feng_restricted_2018}, who proposed a new procedure, Consistent Cross-Validation, a modification on the standard cross-validation procedure.

To our knowledge, no article comprehensively shows that using flexible weights in the weighted bootstrap achieves approximately the same as adopting very different resampling methods such as $k$-fold cross-validation (CV) or non-weight based bootstraps, for example an $m$-out-of-$n$ bootstrap. 
These latter methods have limitations that are often somewhat overlooked. 
Even for methods that can deal with more parameters than observations (large $p$, small $n$) there are minimum requirements on how small $n$ can be. 
For example if the number of observations is close to 10, say, as is often the case in Phase I clinical trials, $k$-fold cross-validation can become infeasible because every design point may be necessary for obtaining valid estimates; reducing the data to $k$ folds simply breaks the estimator.
 Methods that use information from all design points are needed in such and other cases.


In this article, we propose a new method for selecting the Lasso regularization parameter based on a flexible-weighted bootstrap procedure.
An important feature of our proposed procedure is that unlike cross-validation or the usual paired bootstrap, our weighted bootstrap uses every observation in the estimation of the Lasso solution path making it applicable for situations where non-continuous covariates may have low prevalence.
We introduce a new idea of inverting the weights used in the estimation of the solution for the estimation of prediction error which is analogous to the way in which cross-validation and the usual bootstrap estimate prediction error. 
By investigating the performance of different weighting schemes, drawing weights from the Beta distribution due to its wide variety of density function shapes, we show that this parametrization of the weighted bootstrap can be used to emulate among others $k$-fold cross-validation and the $m$-out-of-$n$ bootstrap \cite{shao_bootstrap_1996, mueller_outlier_2005, mueller_robust_2009}. 
In particular, when choosing the parameters to generate a weighting scheme we find a narrow part of the parameter space to perform well in both simulated and real data, where weighting schemes emulating an $m$-out-of-$n$ type bootstrap procedure selects good models on the Lasso path. 
Our methods immediately generalise to any regularization method producing a path through the parameter space but for ease of presentation, and to not mask the main contribution of this article, we solely focus on the most popular shrinkage method.

\section{A weighted bootstrap for the estimation of the Lasso penalty parameter}

Here we outline a pairwise weighted bootstrapping procedure for selecting a suitable regularization parameter, $\lambda$, for the Lasso.
Suppose we have a sample $S = \{(x_i^T,y_i), i = 1,\ldots, n\}$, drawn from a population $P = \{(X_i^T,Y_i), i = 1, \ldots, N\}$, where $N$ is assumed to be very much greater than $n$ and $S$ is a simple random sample such that $S$ is an unbiased representation of $P$. 
Ideally we would like to find a $\lambda$ such that the true "optimal" set of non-zero coefficients are recovered, that is, $\hat{\alpha} (\lambda) = \alpha (\lambda)$.
Since in practice it is not usually possible to obtain further independent samples from the population $P$ of interest, bootstrapping techniques act to emulate the process of obtaining independent samples from $P$ by treating the sampled data as a population from which repeated samples are drawn.
Instead of resampling directly from $S$, the weighted bootstrap samples $n$ weights from a non-negative distribution $W$ and applies these to the original observations to construct a weighted bootstrap sample.
%
%
The resulting weighted bootstrap can then be used to produce a bootstrap estimate $\hat{\bm{\beta}}^*$ of $\hat{\bm{\beta}}_{\mathrm{true}}$, which are the true set of parameter values of interest.
Specifically, the weighted bootstrap estimate for the Lasso for a single bootstrap run with $w_1,\ldots,w_n$ i.i.d. $W$ is defined as 
\begin{equation}
\label{eq3}
\hat{\bm{\beta}}^* (\lambda) = \mathrm{arg\,min}_\beta \sum\limits_{i=1}^n   \left(y_i - \sum\limits_{j=1}^p \beta_j x_{ij}\right)^2 \times w_i  + \lambda \sum\limits_{j=1}^p |\beta_j|.  
\end{equation}
This resampling procedure is repeated $b$ times for some large value $b$ to produce $b$ vectors of parameter estimates $\hat{\bm{\beta}}^*_{1}, \hat{\bm{\beta}}^*_{2} ,\ldots, \hat{\bm{\beta}}^*_{b}$ produced using weight vectors $\bm{w}_1,\ldots,\bm{w}_k$.

If we now define $\bm{u}_k = 1 - \bm{w}_k$ as an `inverse' to the vector of weights $\bm{w}_k$ for the $k^{th}$ bootstrap sample, we can calculate a mean squared prediction error for a particular value of $\lambda$ as
\begin{equation}
\mbox{MSPE} (\lambda) = \sum\limits_{k=1}^b\sum\limits_{i=1}^n (y_i - \sum\limits_{j=1}^p \hat\beta^*_{jk}(\lambda) x_{ij})^2 \times u_{ik}.
\end{equation}
By repeating this process over a range of $\lambda$ values we can choose to use $\lambda^*$, the $\lambda$ which minimizes the MSPE.

\subsection{Choice of weights}

In the following we will describe how $k$-fold cross-validation, the paired bootstrap and the $m$-out-of-$n$ paired bootstrap can be formulated as a weighted bootstrap. In addition to this we will describe a weighting scheme using the Beta distribution which will be the primary focus of our application that has the ability to somewhat mimic these other resampling schemes.

\subsubsection{Beta distribution for weights}

Due in part to the broad variety of weight distributions it can generate, we use the Beta distribution as the primary function for generating weights.
The Beta distribution is a family of continuous probability distributions defined on the closed interval [\,0\,, 1\,].
Parameterised by two shape parameters $a>0$ and $b$ > 0, the probability density function of the
Beta distribution is given by
\begin{equation}
f(z_i; a, b) = \frac{1}{B(a,b)} z^{a - 1}(1-z)^{b -1},
\end{equation}
where $B$ is the Beta function. 

\begin{figure}
\centering
\includegraphics[width=0.60\textwidth]{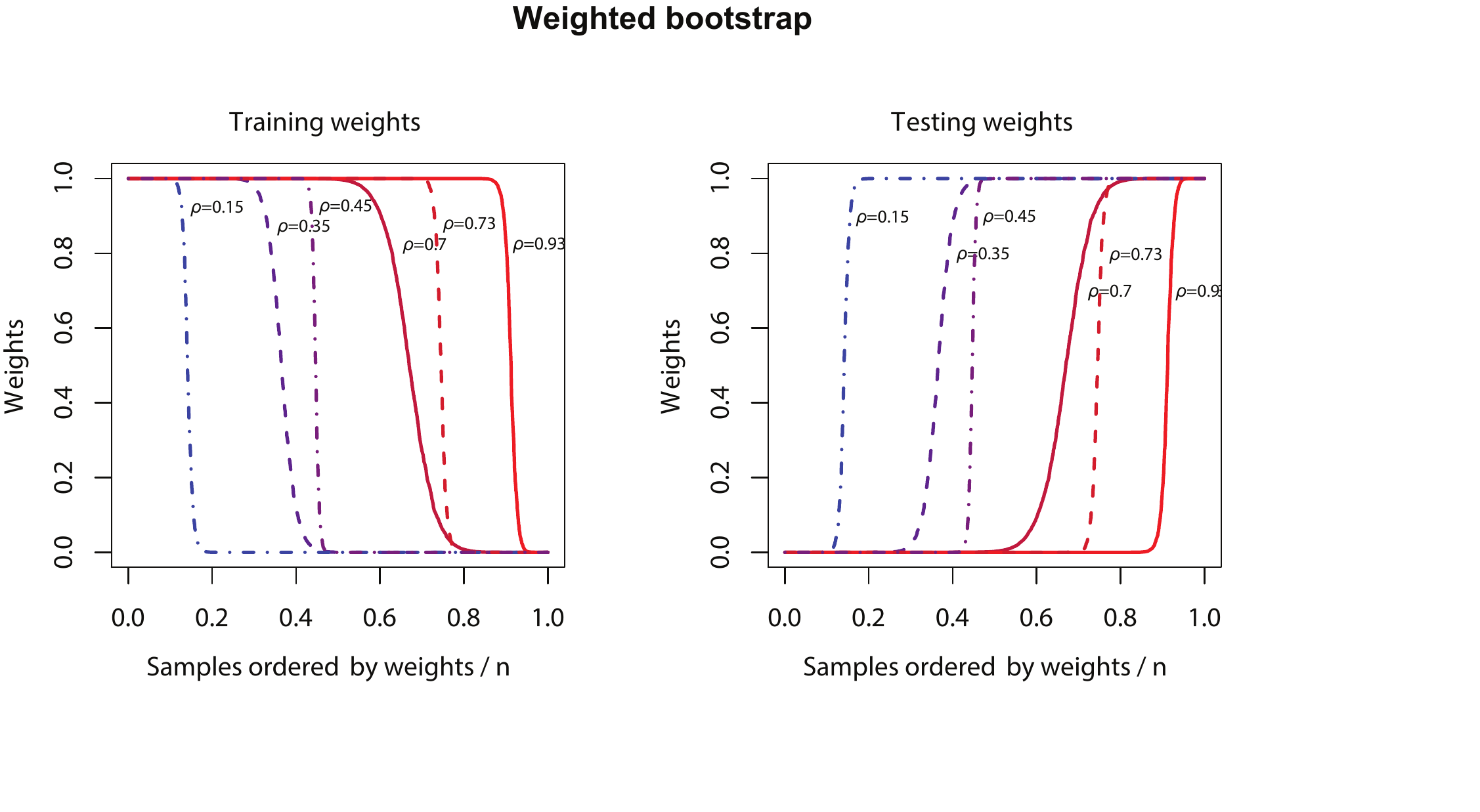}
\caption{Test and training weights for weighted boostrap}
\label{figwb}
\end{figure}

%
The probability density function of the Beta distribution has a large geometry and depending on the values of the shape parameters, can take a wide variety of different shapes. 
Examples of the weights that can be generated by varying the shape parameters of the Beta density function are given in Figure \ref{figwb}.
While it is not made clear by Figure \ref{figwb}, all of these weights are non-zero which is a distinct difference to the cross-validation and bootstrap weights described in the following.

\subsubsection{k-fold cross-validation}

\begin{figure}[h]
\centering
\includegraphics[width=0.60\textwidth]{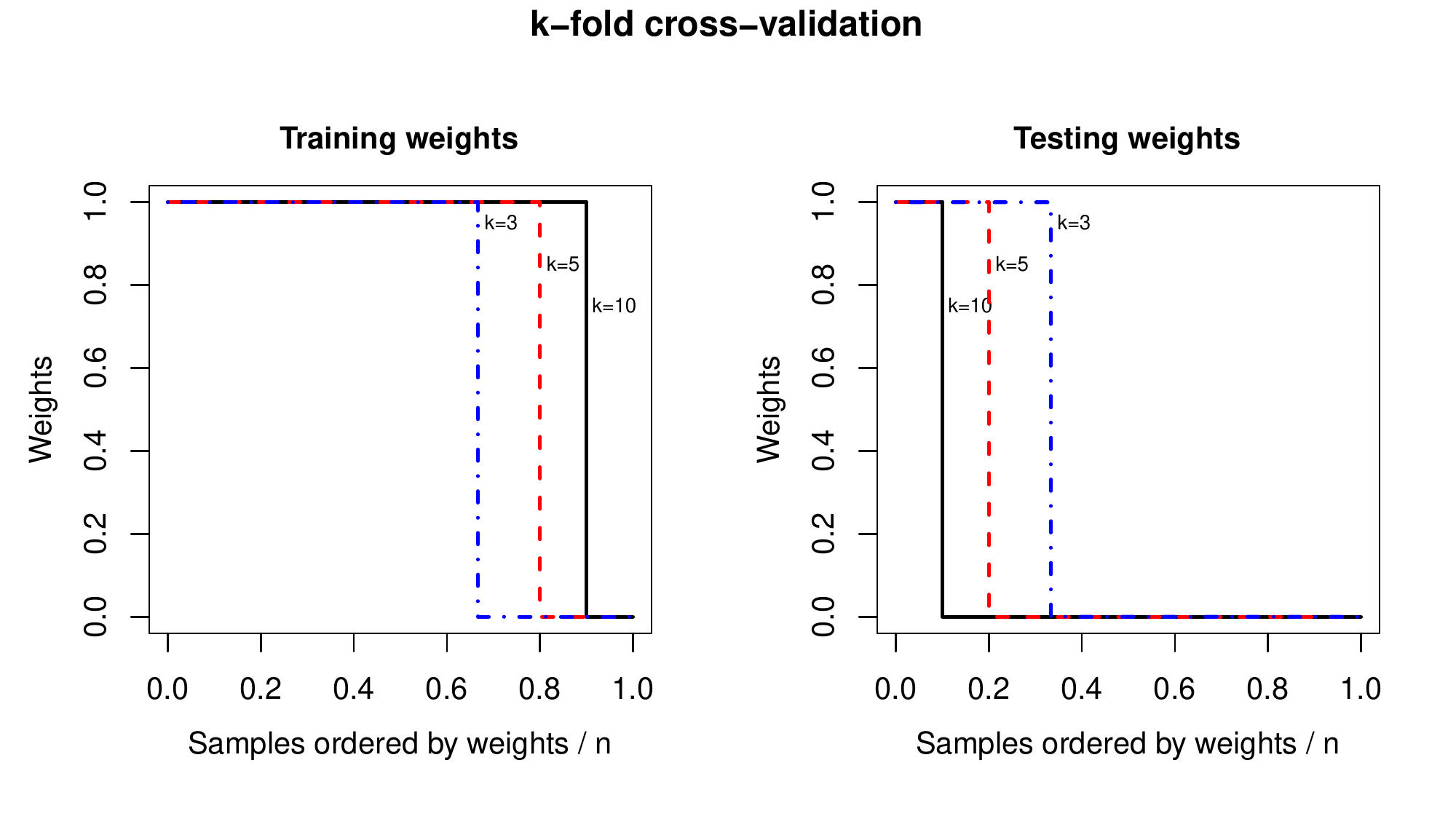}
\caption{Test and training weights for $k$-fold cross-validation}
\label{figcv}
\end{figure}

Perhaps the most commonly used method for selecting the Lasso regularization parameter is $k$-fold cross-validation. 
The key idea of $k$-fold cross-validation is to partition the complete dataset into complementary training and test subsets. 
The Lasso solution is then estimated using the training subset consisting of $(k-1)$ folds and its performance is then validated on the test subset, the $k$th fold, by means of prediction error, iterating through all $k$ folds. 
We can also rephrase $k$-fold cross-validation as a weighted process, where attached to each observation is a training weight $w_i \in \{0,1\}$. 
In the case of 10-fold cross-validation, when sampling without replacement, the first training set can be generated by giving 90\% of the sampled observations a weight of 1 and the remaining 10\% observations a weight of 0.
For any $k$, the vector of weights $\bm{w}$ can be sampled from a multivariate hypergeometric distribution with $h = n \frac{k-1}{k}$
\begin{equation}
p(w_1, w_2,\ldots, w_n) = \frac{ {1 \choose w_1} {1 \choose w_2} \ldots {1 \choose w_n}}{{n \choose h}},
\end{equation}
where $h$ is the number of samples in the training subset. 
Figure \ref{figcv} illustrates the weight structure of $k$-fold cross-validation with $k = 3, 5$ and $10$. 
By comparing the $k$-fold cross-validation weights in Figure \ref{figcv} to a variety of different Beta weights in Figure \ref{figwb}, it can be observed that the Beta weights where $\rho=0.91$ might offer a close approximation to 10-fold cross-validation.

\subsubsection{The paired bootstrap and m-out-of-n paired bootstrap}

As it is generally not possible to obtain further independent samples from a population $P$ of interest, the paired bootstrap emulates the process of obtaining $n$ independent samples from $P$ by drawing $n$ samples with replacement from the original samples, $S$.
In such a set up, sampling process of the paired bootstrap described above corresponds to having multinomial weights $\bm{w}$ \cite{efron_jackknife_1982}, where
\begin{equation}
p(w_1, \ldots, w_n) = \frac{n!}{\prod_{i=1}^n w_i!} \prod_{i=1}^n \left(\frac{1}{n}\right)^{w_i}
\end{equation}
and $\sum\limits_{i=1}^n w_i = n$. It is known that on average, the paired bootstrap puts positive weights on
\begin{equation}
1-\left(1-\frac{1}{n}\right)^n \rightarrow 1 - \frac{1}{e} \approx 0.6322
\end{equation}
of the observations in the original sample \cite{praestgaard_exchangeably_1993}. 
Similar to cross-validation, those observations not drawn in each of the bootstrap samples can be used to obtain approximately unbiased estimates of prediction error \cite{breiman_heuristics_1996}. 
Thus, unlike our other schemes we define the vector of testing weights for the paired bootstrap, $\bm{u}$, as $u_i = 1 - \bm{1}(w_i > 0)$ for $i=1, \ldots , n$.

It has been shown that even for large $n$ the paired bootstrap does not consistently select the true non-zero regression coefficients, that is $P(\hat{\beta} (\lambda) = \beta (\lambda)) \not \rightarrow 1$ as $n \rightarrow \infty$\cite{shao_bootstrap_1996}. 
However, the $m$-out-of-$n$ bootstrap becomes consistent if $m \rightarrow \infty$ and $m/n \rightarrow 0$\cite{shao_bootstrap_1996,mueller_outlier_2005}.
The $m$-out-of-$n$ bootstrap is a generalization of the paired bootstrap where instead of sampling $n$ samples with replacement, $m\neq n$ samples are selected instead where typically $m < n$.
As such the weights for an $m$-out-of-$n$ bootstrap can also be sampled from a multinomial distribution, where
\begin{equation}
p(w_1, \ldots, w_n) = \frac{m!}{\prod_{i=1}^n w_i!} \prod_{i=1}^n \left(\frac{1}{n}\right)^{w_i}
\end{equation}
and $\sum\limits_{i=1}^n w_i = m$.

\begin{figure}
\centering
\includegraphics[width=0.60\textwidth]{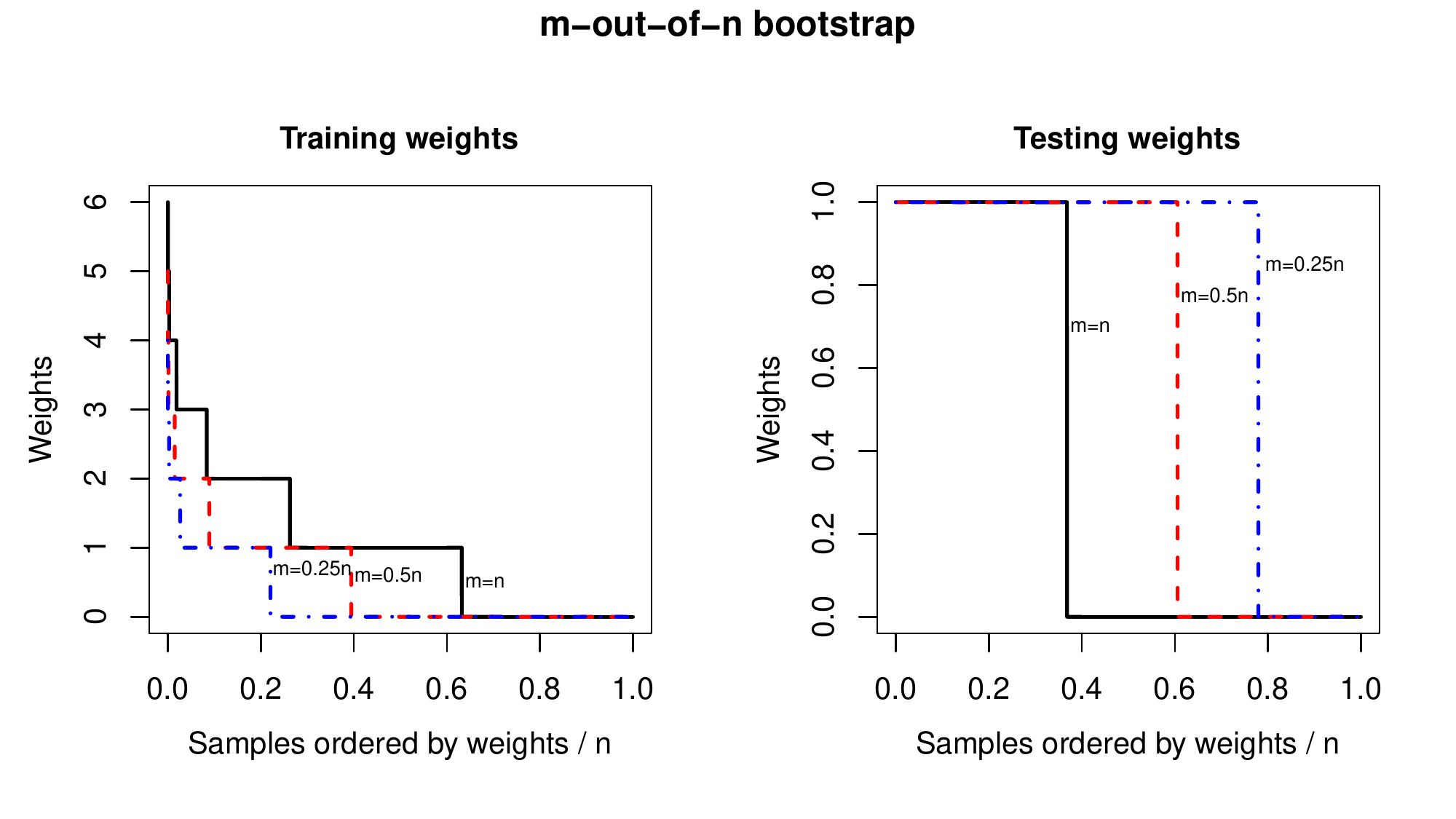}
\caption{Test and training weights for a selection of $m$-out-of-$n$ bootstraps}
\label{figmout}
\end{figure}

Examples of the weights generated for the paired bootstrap and $m$-out-of-$n$ paired bootstrap are given in Figure \ref{figmout}. 
Here it can be seen that the number of samples with non-zero weights is slightly above 60\%.
Also, as expected the number of samples with non-zero weights in the $m$-out-of-$n$ bootstraps decreases as $m$ decreases.
By comparing the bootstrap weights in Figure \ref{figmout} to the Beta weights in Figure \ref{figwb}, it can be observed that the Beta weights where $\rho=0.13$ might offer a close approximation to the $0.25n$-out-of-$n$ bootstrap.

\section{Simulation}

In the following, we explore the behavior of our proposed weighted bootstrap procedure through a simulation. 
The weighted bootstrap will be contrasted with the information criterion EBIC \cite{chen_extended_2008} and the resampling approaches leave-one-out cross-validation (LOOCV), 10-fold cross-validation and the $m$-out-of-$n$ bootstrap.

We base our simulation study on the 'Quadratic Model' version of the Diabetes study that is available in the R package \emph{LARS} \cite{efron_least_2004}. 
This diabetes dataset, $\textbf{X}$, includes $n = 442$ observations, each with $p = 64$ covariates and a response variable, $\textbf{y}$, that is a quantitative measure of disease progression one year after the baseline.
 To construct the parameters for our simulation, a Lasso is used to estimate a model, $\hat{\bm{\beta}}$, from the diabetes study using repeated 10-fold cross-validation to tune the Lasso. 
The $\hat{\bm{\beta}}$ from the fitted model has 15 non-zero coefficients. 
This model is then treated as the truth with corresponding regression parameter vector $\bm{\beta}_{true} = \hat{\bm{\beta}}$, and can be used to calculate the expected value of $\bm{y}$, $\bm{mu_{y}} = \textbf{X}\bm{\beta}$ and the standard deviations of $\bm{y}$, $\bm{\sigma_y}$.
A simulated $\bm{y}$ is then generated from $N(\bm{\mu_y},\bm{\sigma_y^2})$.

\begin{figure}
\centering
 \begin{subfigure}[b]{0.275\textwidth}
\includegraphics[width=\textwidth]{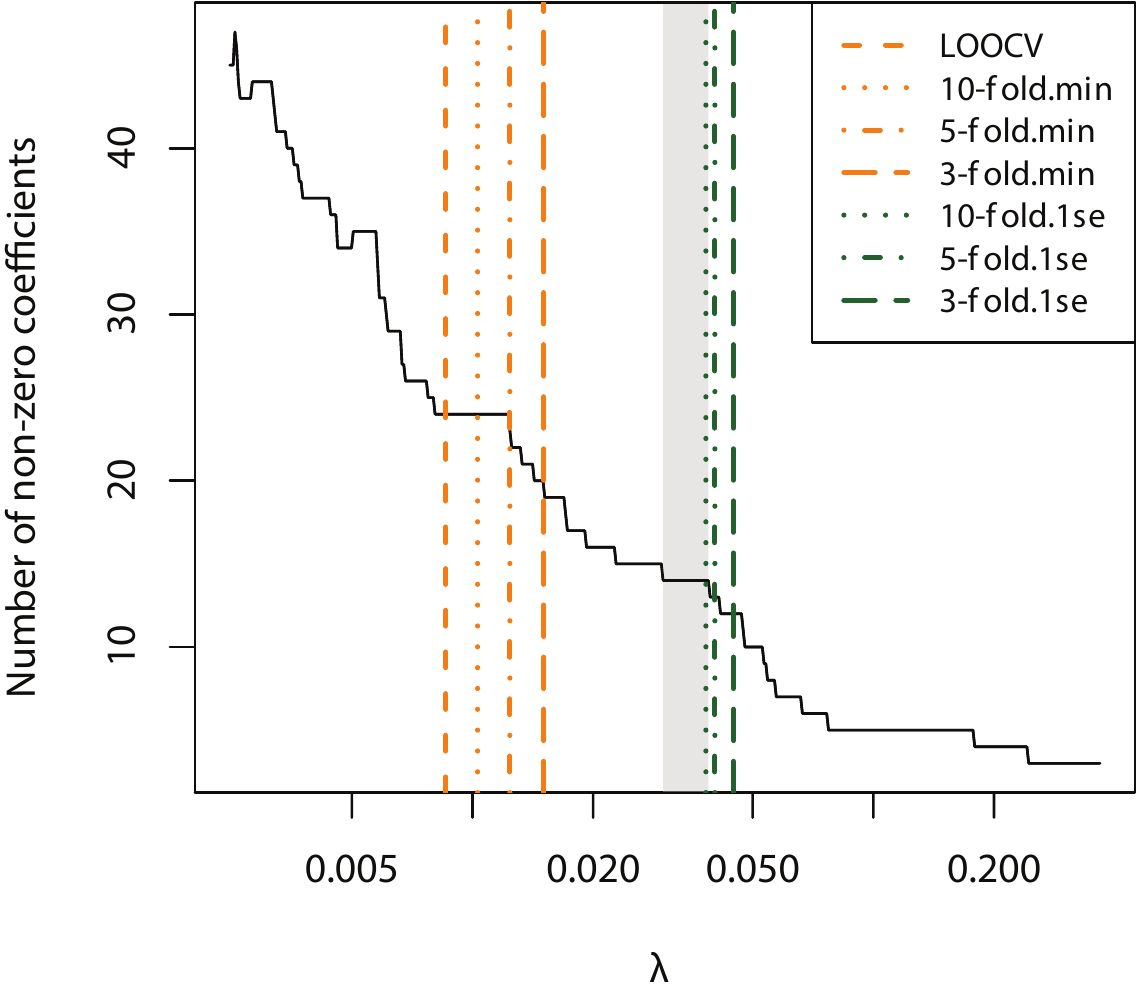}
	\caption{Cross validation}
 \end{subfigure}
 \begin{subfigure}[b]{0.275\textwidth}
\includegraphics[width=\textwidth]{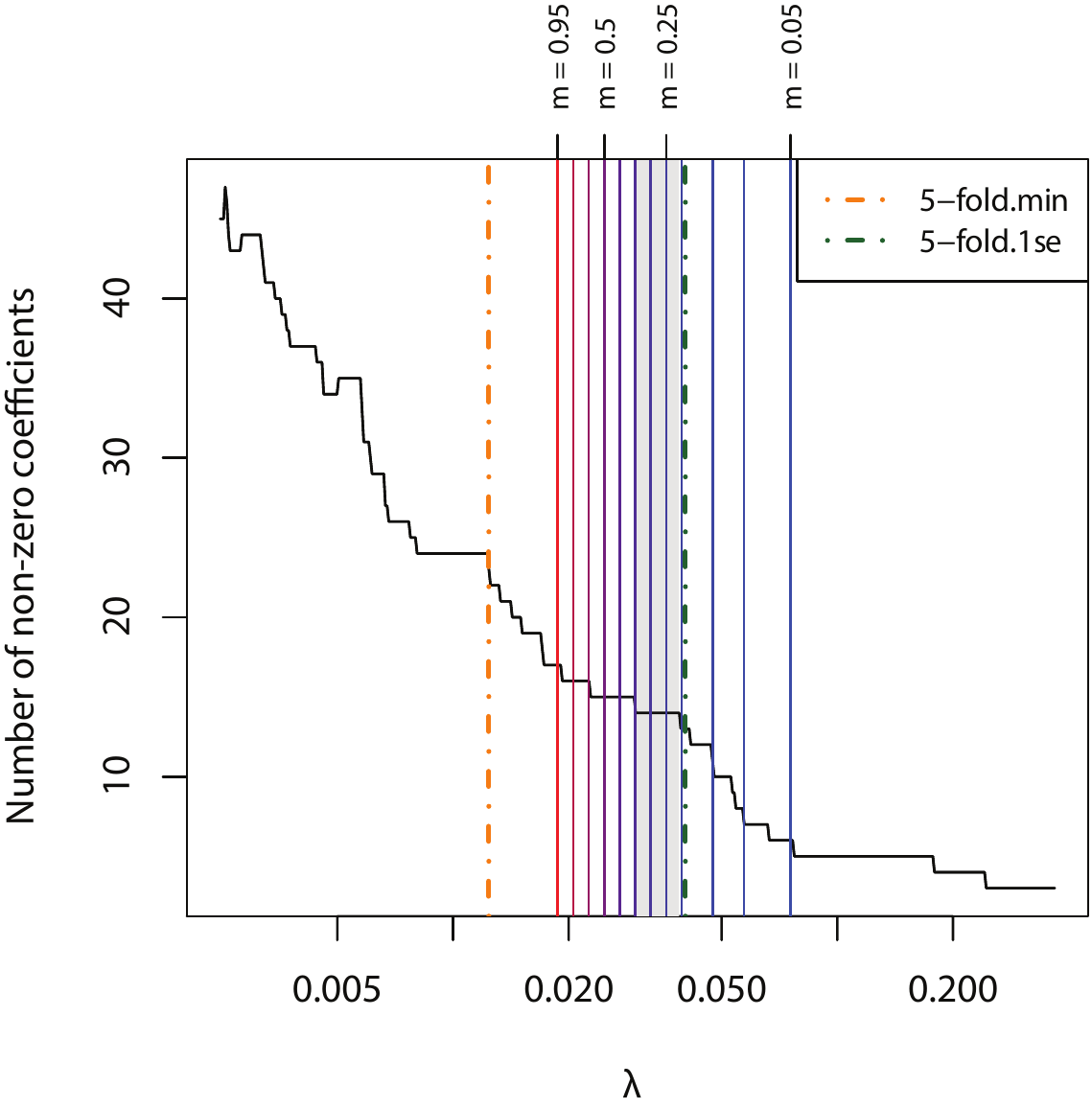}
	\caption{m-out-of-n cross validation}
 \end{subfigure}
 \begin{subfigure}[b]{0.275\textwidth}
\includegraphics[width=\textwidth]{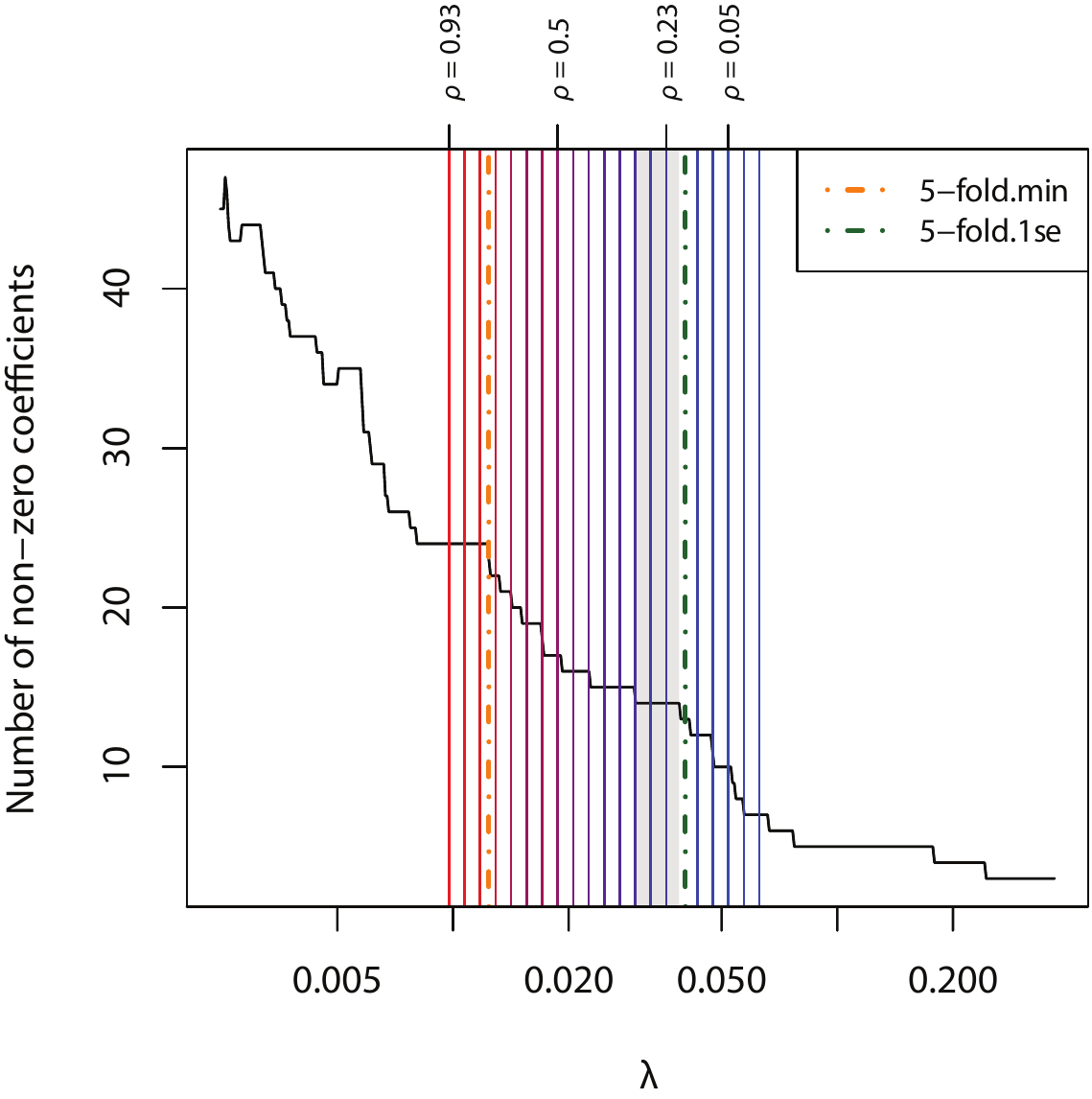}
\caption{Weighted bootstrap}
 \end{subfigure}
\caption{Number of non-zero regression coefficients in diabetes simulation}
\label{figNonZeroDiab}
\end{figure}

\begin{figure}[h]
\centering
 \includegraphics[width=0.3\textwidth]{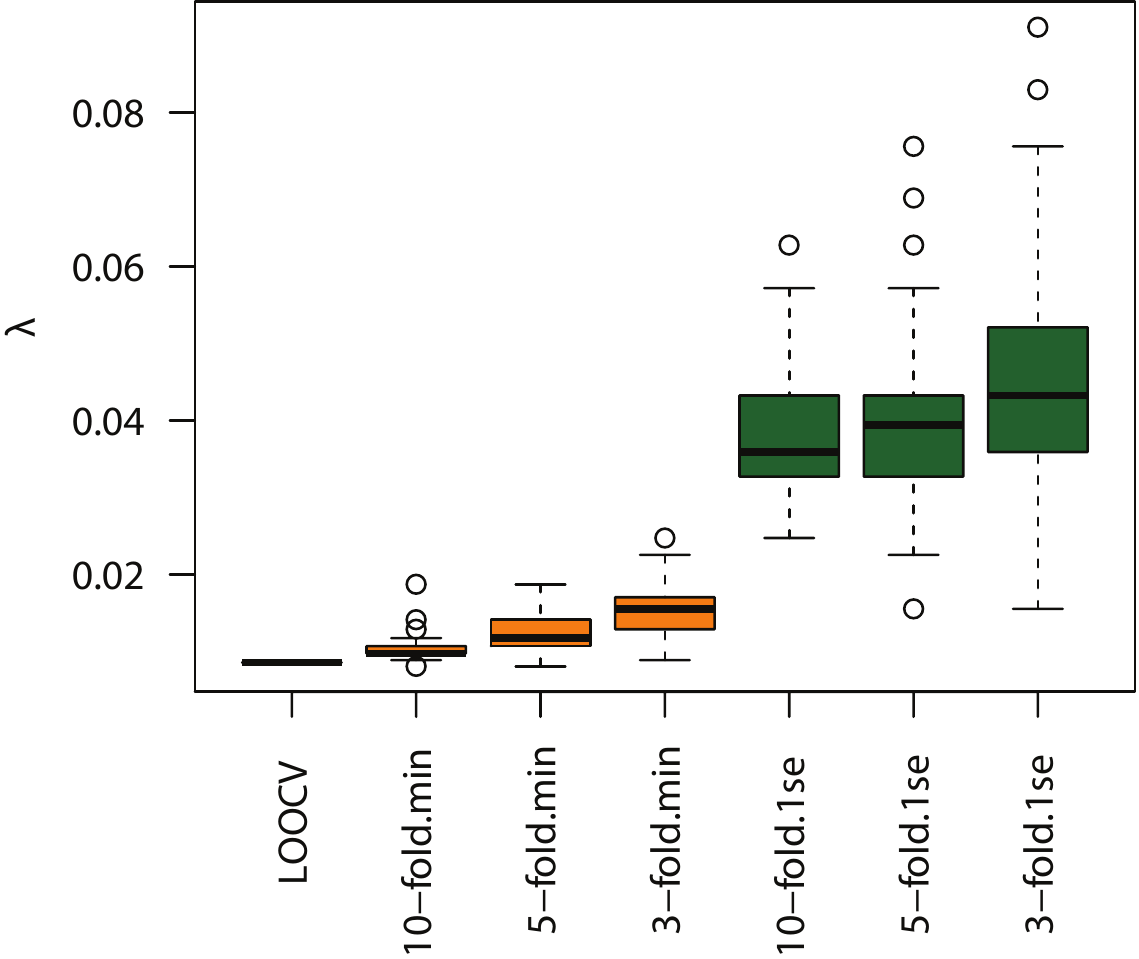}
	\caption{Boxplots of optimal $\lambda$'s}
 \label{figlambox}
\end{figure}

The choice of resampling scheme used to estimate an optimal penalty parameter greatly influences the size of the penalty parameter.
As shown in Figure \ref{figNonZeroDiab}a, as $k$ in the $k$-fold cross-validation decreases the penalty parameter, $\lambda$, correspondingly increases.
The use of LOOCV (i.e. $n$-fold cross-validation) results in the smallest penalty parameter and hence the largest selected model (number of non-zero regression coefficients is 24, Figure \ref{figNonZeroDiab}a).
Conversely, $3$-fold cross-validation gives the largest penalty parameter and the smallest model (number of non-zero regression coefficients is 20, Figure \ref{figNonZeroDiab}a).
This relationship is further illustrated in Figure \ref{figlambox}, where the variance of $\lambda$'s selected also increases as $k$ decreases and the size of the training folds decrease. 
These results illustrate the strong relationship between the size of the training folds and the number of variables included in the final fitted model.

There is also a strong relationship between the choice of $m$ in the $m$-out-of-$n$ bootstrap and the penalty parameter $\lambda$. 
Given that we now expect to see a relationship between the size of the training set and the magnitude of $\lambda$, it is not surprising to see that just like the $k$ in $k$-fold cross-validation, the $m$ from the $m$-out-of-$n$ also shares a monotonic relationship with $\lambda$. 
When $m$ is chosen to equal one, approximately two thirds of the data would be used in the training set.
In Figures \ref{figNonZeroDiab}a and \ref{figNonZeroDiab}b the $m$-out-of-$n$ bootstrap with $m$ close to $n$ produces a very similar optimal $\lambda$ as in 3-fold cross-validation. 
This observation is important for emphasising the relationship between the size of the training set and selection of the penalty parameter $\lambda$.

\begin{figure}
\centering
 \begin{subfigure}[b]{0.3\textwidth}
\includegraphics[width=\textwidth]{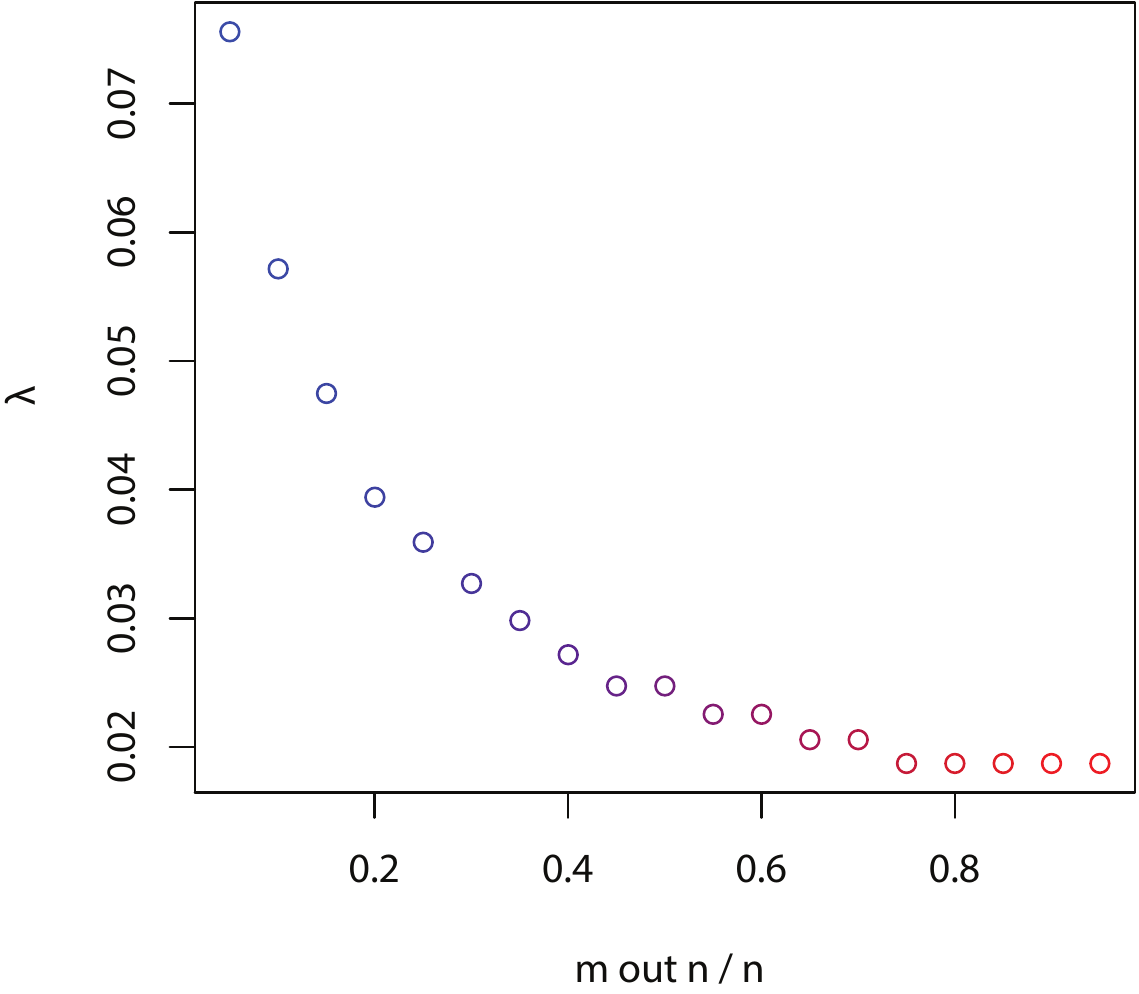}
	\caption{m-out-of-n cross validation}
 \end{subfigure}
 \begin{subfigure}[b]{0.3\textwidth}
\includegraphics[width=\textwidth]{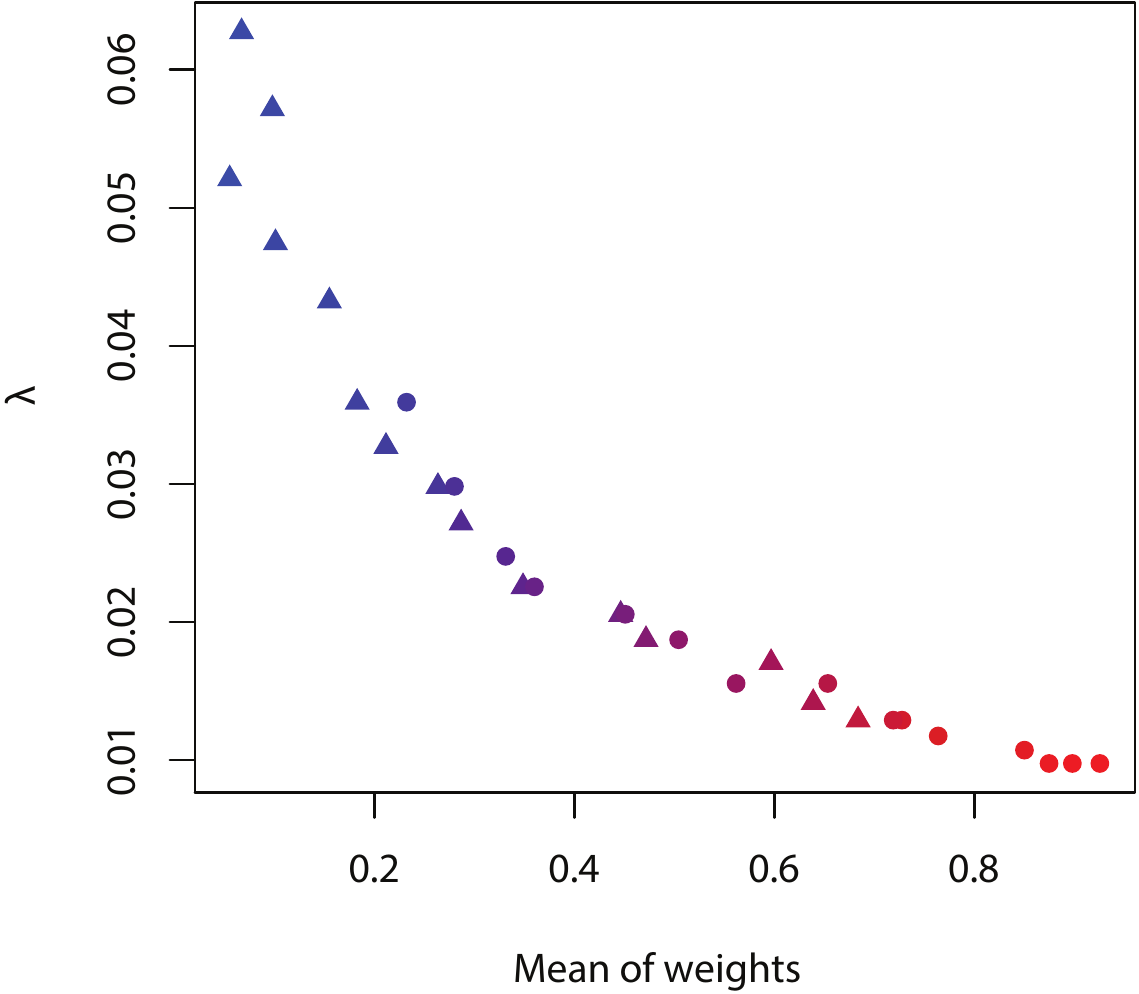}
\caption{Weighted bootstrap}
 \end{subfigure}
\caption{Lambda plots diabetes}
\label{figLambdaDiab}
\end{figure}

The parameterisation of the weights used for the weighted bootstrap are not directly important for interpreting the relationship between the choice of weights and the selection of $\lambda$.
To illustrate the behaviour of the weighted bootstrap we chose to use a Beta distribution with varying selections of the two shape parameters $a$ and $b$ to construct the weights. 
To demonstrate the relationship between these parameters and choice of $\lambda$, consider a new parameter $\rho$ which represents the area under the curve defined by the ordered weights in Figure \ref{figwb}.
This new parameter $\rho$ is then just the average of the observed weights and is representative of the proportion of samples used in the training set of the weighted bootstrap.
In Figure \ref{figNonZeroDiab}c there is a monotonic relationship between $\rho$, the proportion of samples used in the training set, and the choice of $\lambda$.
Additionally, the strong relationship between $\rho$ and $\lambda$ is minimally affected by choices of shape parameters and $a$ and $b$ as observed in Figure \ref{figLambdaDiab}.
The parametrization of $\rho$ as the average of the observed weights and its monotonic relationship with $\lambda$ will be very valuable when interpreting the weighting schemes which select optimal $\lambda$.

\begin{figure}[b]
\centering
 \begin{subfigure}[b]{0.275\textwidth}
\includegraphics[width=\textwidth]{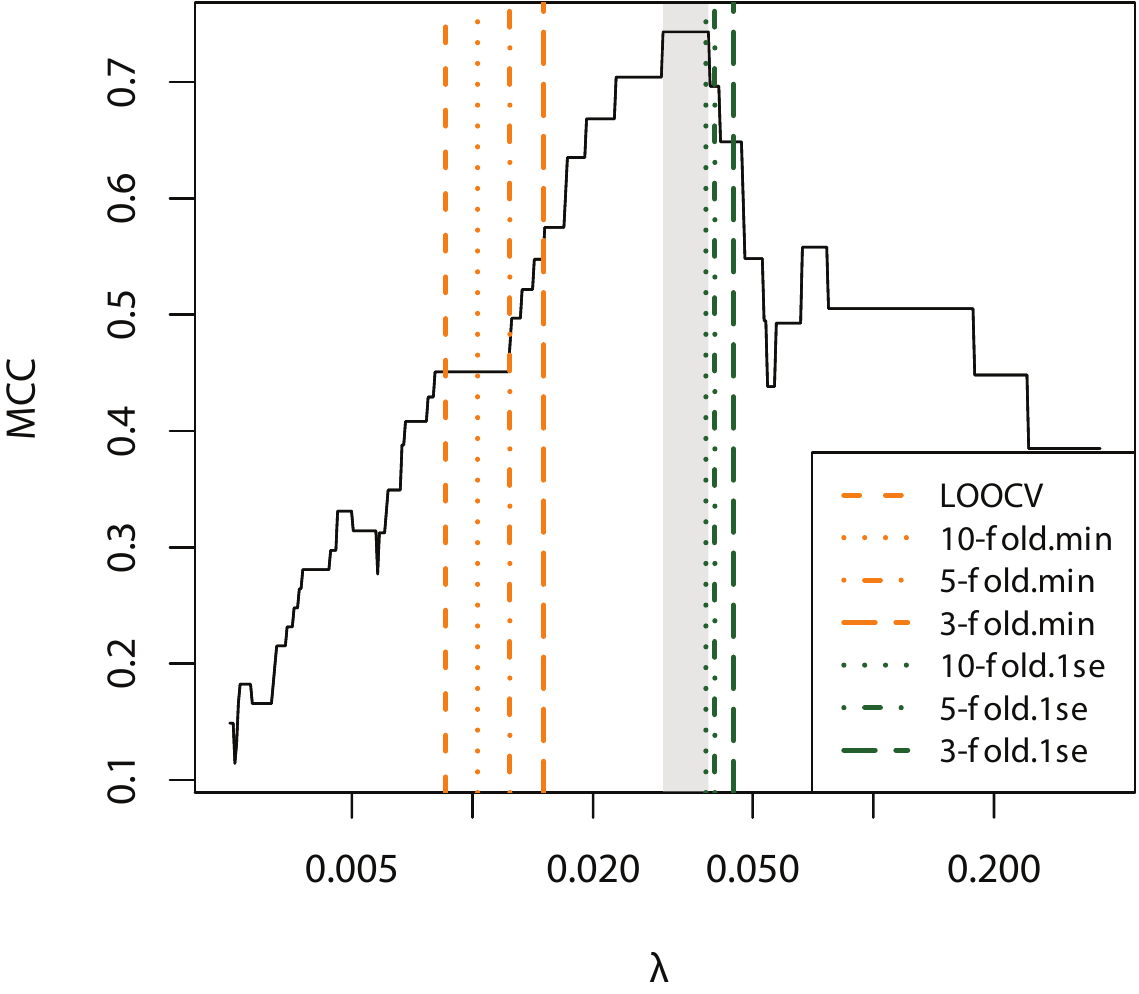}
	\caption{Cross validation}
 \end{subfigure}
 \begin{subfigure}[b]{0.275\textwidth}
\includegraphics[width=\textwidth]{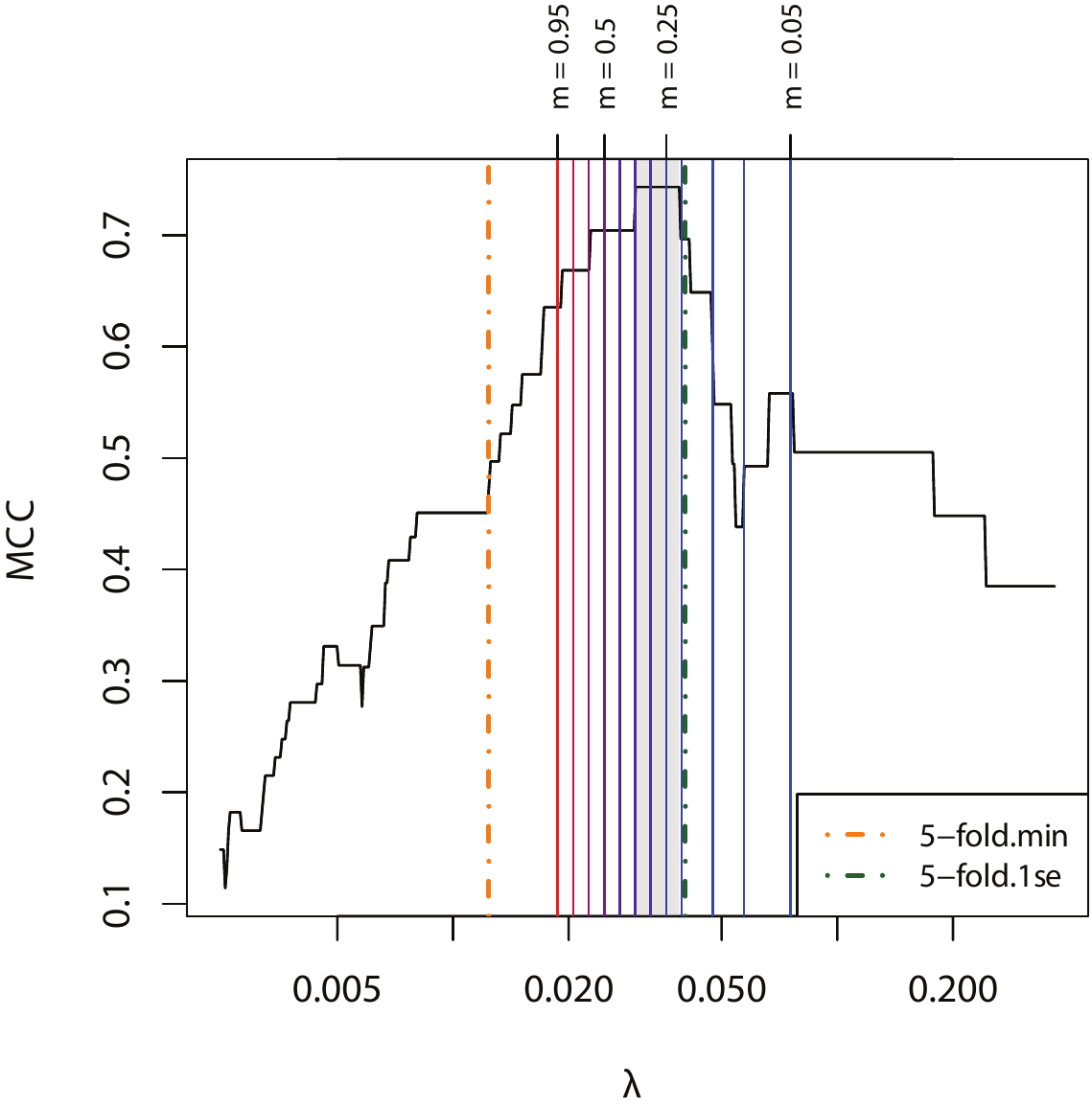}
	\caption{m-out-of-n cross validation}
 \end{subfigure}
 \begin{subfigure}[b]{0.275\textwidth}
\includegraphics[width=\textwidth]{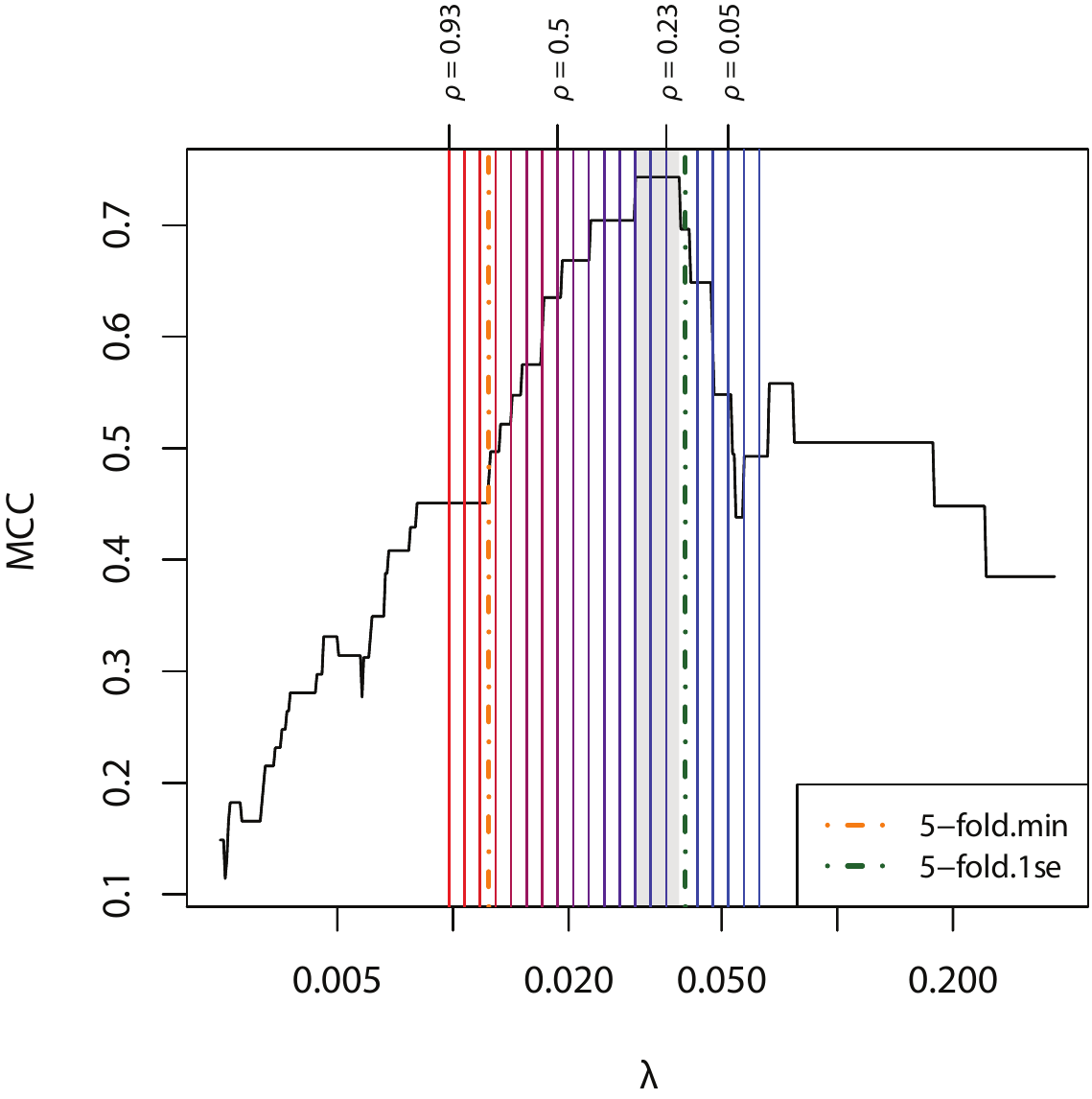}
\caption{Weighted bootstrap}
 \end{subfigure}
\caption{MCC plots from diabetes simulation}
\label{figMCCDiab}
\end{figure}

In our simulation, $k$-fold cross-validation does not select an ideal penalty parameter.
Every $k$-fold cross-validation scheme evaluated that selects penalty parameters by minimising the cross-validation error rates, selected penalty parameters that produce models that are larger than the true model and hence include non-informative variables.
Figure \ref{figMCCDiab}a compares the relationship between the Matthew's correlation coefficient (MCC) and penalty parameter for the full solution path of the Lasso. 
The MCC functions as a balanced measure to assess the correlations between true and false positives and negatives of a predicted model.
The MCC ranges between negative one and positive one, with a value of one being ideal.
No model in the solution path of the Lasso recovers the true model when applied to the simulated data, that is, the MCC is not equal to one.
However, there is a span of penalty parameters that have the largest MCC and produce models with 14 coefficients.
No $k$-fold cross-validation selects penalty parameters that lie in the range with largest MCC.

When applied to $k$-fold cross-validation, the "one-standard-error" rule does select models that have the largest MCC.
The one-standard-error rule has become a popular heuristic for selecting parsimonious models \cite{hastie_elements_2009}.
This heuristic selects larger penalty parameters, selecting the penalty parameter that has a prediction error that is one standard error away from the minimum prediction error.
Figure \ref{figMCCDiab}a shows that the $\lambda$ selected by this rule are all larger for 10-fold, 5-fold and 3-fold cross-validation and lie in or near the region with largest MCC.
The behavior of the one-standard-error rule in Figure \ref{figNonZeroDiab}a displays the impact that the one-standard-error rule can have on the complexity of a selected model, with the one-standard-error rule selecting models that are half the size of the standard minimum-error approach.


The weighted bootstrap can facilitate the interpretation of model selection heuristics. 
While the one-standard-error rule is regularly used in practice, there exists very little intuition as to the properties of its selected models.
The only certainty is that penalty parameters selected using the one-standard-error rule will generate sparser models than penalty parameters selected by minimizing the $k$-fold cross-validation prediction error.
As the behaviour of the weighted bootstrap is dependent on $\rho$, the proportion
of samples used in the training set of the weighted bootstrap, this provides a clear path to interpreting the penalty parameter selections of other heuristics.
In Figure \ref{figMCCDiab} the penalty parameter chosen with the one-standard-error rule is equivalent to one selected by weighted bootstrap with $\rho$ slightly smaller than 0.23. 
So in this simulation, using the one-standard-error rule is equivalent to using a training set with approximately $23\%$ of the data.
In practice, it is clear that the weighted bootstrap can be used to provide context to the behaviour and performance of the one-standard-error rule.

\section{Application to the integration of micro-RNA and phenotypic data}

MicroRNA (miRNA) are a class of small RNA molecules, approximately 21 nucleotides long, that regulate the expression of messenger RNA (mRNA). 
It is mRNA that is typically measured in a transcriptomic or gene expression study.
While there are thought to be over 20,000 genes in the human genome, there are only approximately 1000 known miRNA.
With a single miRNA regulating many mRNA, miRNA provide a natural dimension reduction of the human genome improving the robustness of studies associating the expression of genes with a disease related outcome.

Aberrant expression of miRNA and their target mRNA have been implicated in several neurodegenerative disorders including Alzheimer's disease \cite{patrick_dissecting_2017}.
The Religious Orders Study (ROS) and Memory and Aging Project (MAP) are two longitudinal studies of aging that are designed to be complementary \cite{bennett_neuropathology_2006}.
There are 709 subjects in ROSMAP that have measurements of 309 miRNA.
Accompanying the miRNA measurements are 38 clinical, pathological and genetic variables (clinical variables), including a measurement of cognitive decline.
In the following we will use the miRNA and clinical variables to model cognitive decline.

Penalised regression can be used to model cognitive decline with miRNA expression and other clinical variables.
Given the scale of gene expression studies, there has been much difficulty integrating clinical data with gene expression data in predictive models without the signal in the gene expression data swamping that of the clinical data\cite{tibshirani_pre-validation_2002}.
As there are only 309 miRNA expressed in the ROSMAP dataset, it becomes plausible to directly concatenate these miRNA with the set of 39 clinical variables to use as covariates in a model.
Still, with 348 covariates feature selection is warranted.
Using Lasso with 5-fold cross-validation and the one-standard-error rule produces a model of cognitive decline which includes two pathological variables, \textit{neurofibrillary tangles} and \textit{tau burdan}, a genetic variable, \textit{ApoE4}, and two miRNA, \textit{miRNA-129} and \textit{miRNA-132};
\begin{equation}
\mbox{Cognitive decline} = -0.11 \ - \ 0.0043 \ \mbox{ApoE4} \ - \ 0.0056 \ \mbox{NFT} \ -0.08 \ \mbox{TAU} 
	    \ + \ 0.028 \ \mbox{miRNA-129} \ + \ 0.034 \ \mbox{miRNA-132.}
\end{equation}
This model is consistent with what is known about loss of cognition given that \textit{ApoE4} is the major genetic risk factor for Alzheimer's data \cite{kim_role_2009}, \textit{neurofibrillary tangles} and \textit{tau burdan} are the key pathologies that are associated with neuronal degredation \cite{bennett_neuropathology_2006} and \textit{miRNA-129} and \textit{miRNA-132} are two miRNA which are consistently associated with Alzheimer's disease\cite{patrick_dissecting_2017}.

\begin{figure}
\centering
\includegraphics[width=.3\textwidth]{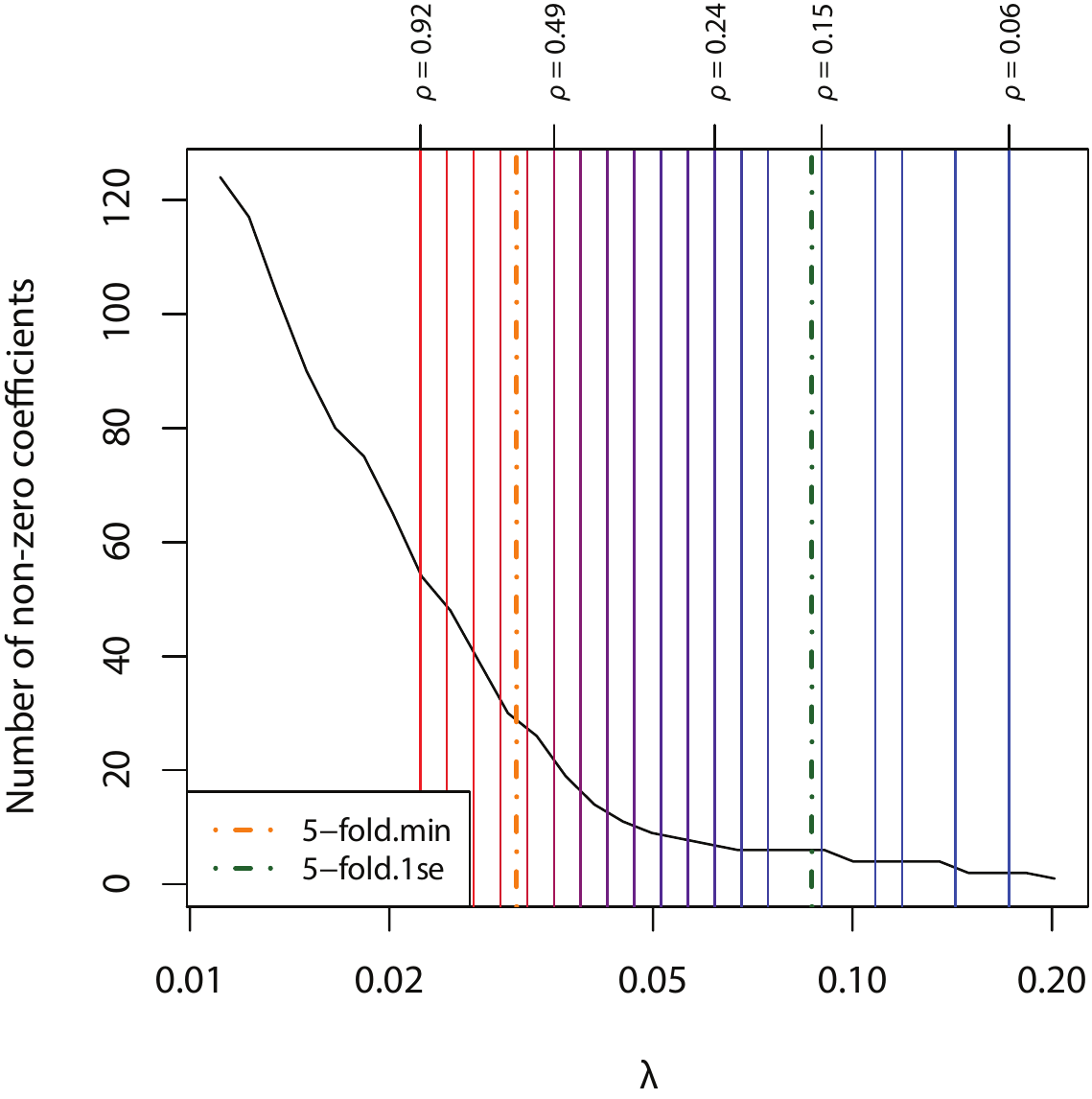}
\caption{Number of non-zero coefficients in miRNA data}
\label{figNonZeromiRNA}
\end{figure}

The weighted bootstrap is useful for evaluating the construction of the model of cognitive decline.
As with the earlier simulations, when a spectrum of weights are used, there is a clear relationship between the proportion of samples used in the training set of the weighted bootstrap, $\rho$, and the Lasso penalty parameter $\lambda$ (Figure \ref{figNonZeromiRNA}).
The $\lambda$ that is selected by the one-standard-error rule is equivalent to the $\lambda$ selected when only $15\%$ of the samples are used to train a model.
This provides two possible interpretations, first the signal from the variables included in model is likely to be strong enough to be present when only $15\%$ of the samples are used to train a model.
Second, models that are trained on a larger proportion of samples are highly variable which is unsurprising given the ratio of $n$ to $p$ and the strong correlation structure present in the miRNA data.  
Regardless, without the use of the weighted bootstrap it would not be possible to hypothesize about an apparently arbitrary penalty parameter chosen by a heuristic akin to the one-standard-error rule.

\section{Discussion}

We have proposed the weighted bootstrap as a strategy for selecting an optimal penalty parameter for a Lasso regression model.
The weighted bootstrap is both an alternative and complementary to the widely used approach of cross-validation.
The performance and behaviour of the weighted bootstrap strategy is demonstrated via simulation and application to the integration of clinical and microRNA data.
These evaluations demonstrate that the weighted bootstrap has significant value when used to find an optimal penalty parameter.

The weighted bootstrap can be used to increase interpretation of penalty parameter selection.
It is well established that the choice of the number of folds used for cross-validation, $k$, will influence the optimised value of penalty parameter $\lambda$ with smaller $k$ increasing $\lambda$\cite{zhang_cross-validation_2015}.
Acknowledging this relationship is valuable as it provides a layer of interpretation to the value of $\lambda$ by translating the choice of a parameter with uninterpretable scale into a choice of $k$ which has an interpretable context.
This benefit holds for both, $m$ from the $m$-out-of-$n$ bootstrap and for $\rho$ from the weighted bootstrap.
As the weighted bootstrap is a generalisation of cross-validation and the $m$-out-of-$n$ bootstrap, it can facilitate the translation of $\lambda$ across the entire range of $\rho$, the proportion of samples used in the training set.

The weighted bootstrap makes it possible to include binary variables with low prevalence in a model.
Variables that describe rare events, like genetic mutations, often cannot be included in modelling when cross-validation with small $k$ is used.
Without using a heuristic, if a binary variable has prevalence smaller than $k$ it cannot be considered as there will be training sets of size $k-1$ which do not include the event and so will be collinear with the intercept.
The weighted bootstrap overcomes this limitation by using all variables in the training set.

In summary, the weighted bootstrap is a powerful tool that can be used for model selection alone, or in combination with other approaches. Specifically, the weighted bootstrap can be used to add a layer of interpretability to penalty parameter selection for regularized regression. It has particular practical value when the sample size $n$ is small to moderate or whenever resampling as occurring in cross-validation or the paired bootstrap does not work with non-continuous explanatory variables.

\subsection{Acknowledgements} This research was supported by the Australian Research Council discovery project grant DP180100836. Aspects of this article were first explored in a honours thesis by Diane Loo under the supervision of one the authors. Diane was also involved in the initial planning of this article.

\nocite{*}
\bibliographystyle{plain}



\end{document}